\documentstyle[aps,prl,psfig]{revtex} 
\author{A. Harnos$^{1}$, G. Horv\'ath$^{1}$, A. B. Lawrence$^{2}$ and G. 
Vattay$^{3}$\\ 
$^1$ University of Veterinary Science, H-1078 Budapest, Istv\'an u. 2,
Hungary\\ 
$^2$ Genetics and Behavioural Sciences Department, The Scottish 
Agricultural College, \\
Bush Estate Penicuik, EH26 0QE, UK\\ 
$^3$ E\"otv\"os University, Department of Physics of 
Complex Systems, \\
H-1117 Budapest, P\'azm\'any P\'eter s\'et\'any 1/A, Hungary}
\title{Scaling and Intermittency in Animal
Behavior }
\begin{document} 
\maketitle
\begin{abstract}
Scale-invariant spatial or temporal patterns and L\'evy flight
motion have been observed in 
a large variety of biological systems. It has been argued that 
animals in general might perform L\'evy 
flight motion with power law distribution of times between two changes 
of the direction of motion. 
Here we study the temporal behaviour of nesting gilts.  
The time spent by a gilt in a given form of activity has power law  
probability distribution without finite average. Further analysis 
reveals intermittent eruption of certain periodic behavioural sequences
which are  responsible for the scaling behaviour and indicates the existence of 
a critical state. We show that this behaviour is in close analogy
with  temporal sequences of velocity found in turbulent flows,  
where random and regular sequences alternate and form an 
intermittent sequence.
\end{abstract}

Scale-invariant spatial and temporal patterns have been observed in 
a large variety of biological systems\cite{1}. It has been demonstrated
that ants\cite{2}, Drosohyla\cite{3} and the wandering albatross, Diomedea 
exulants\cite{4} perform motion with power law distribution of times
between two changes of the direction of motion. The power law
distribution of times then leads to an anomalous L\'evy type diffusion
in space. In the last few years an increasing interest
has been devoted to these superdiffusive processes in 
physics\cite{levi1,levi2,levi3}. 
and in econophysics\cite{econo1,econo2,econo3,econo4}.
Inspite of the extensive experimental studies the detailed mechanism 
responsible for the creation of the underlying power law 
distributions is not well understood. 
In this letter we demonstrate the first time that the power law and 
scaling observed in the behaviour of certain animals is related to 
intermittency, a phenomenon familiar from the theory of dynamical 
systems and turbulence. 

It is well known that non-hyperbolic dynamical 
systems show superdiffusive behaviour. In dissipative systems it is 
caused by the trapping of trajectories in the neighborhood of
marginally unstable periodic orbits\cite{geisel}. 
The paradigmatic system showing such behaviour is Manneville's
one dimensional map\cite{Manneville}
\begin{equation}
x_{n+1}=x_n+cx_n^z (mod 1),
\end{equation}
where $z\geq 1$. Here the sequence $x_n$ spends long time trapped in the
neighborhood of the marginally unstable periodic orbit (a fixed point) 
$x=0$. In analytic maps typically $z=2$. The invatiant density
behaves as $\varrho(x)\sim x^{1-z}$ near the origin and it is not 
normalizable for $z\geq 2$.  Accordingly the distribution of times spent by the
sequence near the unstable periodic orbit has a power law tail.

Next we will show
experimental evidence on the existence of unstable periodic
patterns in animal behaviour. 

Members of the species {\em Sus scrofa} invest considerable time 
and effort into building a nest before farrowing. Our aim 
was to investigate the temporal pattern of this highly motivated 
activity. Using time-lapse video we recorded the behaviour of 
27 gilts and analyzed the last 24 hours preceding 
the farrowing. The experimental subjects were Large White X Landrace 
gilts (Cotswold Pig Development Company, Lincoln, UK). On day 109 
of pregnancy they were moved to their individual farrowing 
accommodations. 
A behavioural collection program\cite{Key} was run to take data from 
the tapes. The behaviour of gilts was classified into eight mutually 
exclusive categories (see Table I). Further details of the experiment 
are published in Ref.\cite{Horvath1}.  
 
As a first step we assigned a symbol 0,1,...,6 to the 7 different types 
of behaviour listed on Table I. The records of the 27 gilts contain 
approximately 24.000 symbols. Then we computed the probabilities of 
the occurrences of symbolic sequences formed from the symbols. This has 
been done by evaluating the decimal value of the base seven number 
coded by the sequence. For example the code sequence 0156 is evaluated 
to be $\underline{0}*1+\underline{1}*7+\underline{5}*7^2+ 
\underline{6}*7^3=2310$.  
 
On the histograms of Fig. 1a, 1b and 1c we 
show the probabilities of the length 3, 4 and 5 sequences respectively. 
One can see that certain symbol sequences occur with high probability 
which does not decrease with increasing symbol length significantly, 
while the majority of symbols have relatively low probability and the 
occurrence of each particular symbol decreases with increasing length. 
 
On the histogram of Fig. 2 we ordered the symbols of lengths $L=2,3$ 
and $4$ according to decreasing probability. One can see that the  
tail of the histogram is exponential. An exponential histogram of 
probability ordered symbols is a property of random texts  
and reflects the fact that most of the behavioural patterns are 
generated by a Markovian process in a purely stochastic way.  
On the other hand, the most probable part cannot be considered 
as a result of random uncorrelated processes. 
On the histogram we can see 
that the probability of the most probable sequences does not 
decay significantly with increasing length and stays approximately 
constant for $L=2,3$ and $4$. 
 
We have identified these most probable behavioural sequences. 
For  gilts kept in pens the most likely sequence is the 
cyclic repetition of "nosing floor"-"alert"-"nosing floor"-...  
pattern. 
This sequence is the consequence of the nest building instinct 
which governs the behaviour of the animal most of the time. 
In crates however there is no possibility to try to build a nest  
due to the limited space and the absence of straw. Gilts become 
frustrated and the ideal "nosing floor"-"alert"-"nosing floor"-... 
sequence is occasionally interrupted by periods of rest. 
On Fig. 3 we show a typical record from a gilt kept in pen. 
We can see that the deterministic sequences of 1-6-1-... 
are dominant interrupted eventually by short random-like 
sequences of other symbols. This is very reminiscent of  
temporal sequences of velocity found in turbulent flows,  
where random and regular sequences alternate and form an 
intermittent sequence. 
 
We can quantify this qualitative analogy by studying the  
probability of periodic sequences. On Figure 4a we show 
the probability to observe the periodic sequences 
"4-6-4-6-..." and "1-4-1-4-..." as a function of the length $L$ 
of the sequence. These are typical periodic symbolic sequences.  
To add a new symbol to an existing sequence in this case is
approximately an uncorrelated, Markovian process. 
The probability decays exponentially with the length.
On the other hand, the probability to observe the special 
periodic sequence "1-6-1-6-..." decays very slowly, according  
to the power law $1/L^2$ for a wide range of $L$ values until
an upper cutoff $L\approx 30$ is reached (Fig. 4b). This is in 
close analogy with intermittent 
flows, where the probability of regular velocity
patterns decays according to a power law.   
The occurrence of the correlated sequences has a drastic effect 
on the probability distribution of the time the animal 
spends engaged in a given type of behaviour as we show next. 
On Fig. 5 we show the probability distribution of these times. 
The data can be fitted very accurately with the power law 
\begin{equation} 
P(t)=C\frac{1}{(t+t_0)^2},\label{dis} 
\end{equation} 
where $t_0=21.3\pm0.6 $ sec. The exponent $2$ of this power law
is in accordance with the similar power law found for the
probability of the "1-6-1-6-..." sequences. 
This function is valid between some lower 
$t_l$ and upper $t_u$ cutoff times. Based on the available data 
we have not reached these and we can say that $t_l$ is less than 30 
seconds and $t_u$ is more than 2000 seconds.  
The distribution (\ref{dis}) is normalizable, however it has no first and 
higher moments. For example the average time spent  in an activity 
\begin{equation} 
\bar{t}=\int dt t P(t) 
\end{equation}  
does not exist. Taking into account that the validity of (\ref{dis}) 
is limited between lower and upper cutoffs the average time spent 
in an activity becomes $\bar{t}\sim t_0\ln(t_u/t_l)$, which is a 
very slowly growing function of $t_u$. However the variance  
of the time spent in an activity $\bar{(\Delta t)^2}\sim t_0t_u$ 
and higher moments are very large making the behaviour of the animal 
very unpredictable. 

It is important to note, that the observed power law distributions 
above are the same as those observable in the Manneville system\cite{Manneville,Ben,geisel}
for $z=2$. This indicates that the sequence "1-6-1-6-..." marks
a marginally unstable periodic orbit of the dynamics.

As a summary we demonstrated here, that the scaling behaviour in the
animal behaviour found here is not 
related to environmental factors like in case of foraging animals, 
where the distribution of food might be distributed in a complex way 
forcing animals to follow L\'evy flight patterns. The environment of 
gilts in pens and crates is almost absolutely unmotivating. The source 
of this complex behaviour can come only from the neural system 
forced by hormonal stimulus due to nesting instincts.  
This is the first carefully examined case, where complex 
scaling behaviour of animals is related to the self-organization 
and possibly to some unstable critical state of the nervous system. 
We hope, that further investigations of the behaviour of other types of 
animals subject to some internal hormonal pressure can be investigated 
and the existence of such a critical state can be established.   

This work has been supported by the Hungarian Science Foundation
OTKA (F17166/T17493/T25866) and the Hungarian Ministry of Education.

\pagebreak 
\begin{table}
\centering
\begin{tabular}{|c|l|l|}        \hline

Symbol & Behaviour & Definition \\ \hline \hline
0 & Comfort & scratching \\ \hline
1 & Alert               & all other behaviours \\
  &                     & not defined here  \\ \hline
2 & Nosing environment  & exploratory or manipulative \\
  &                     & behaviour directed at fixed \\
  &                     & features of the environment \\
  &                     & above floor level \\ \hline
3 & Excretion           & voiding of faeces or urine \\ \hline
4 & Rest                & lateral lying with the head \\
  &                     & lowered and not 'nosing'  \\ \hline
5 & Feeding or drinking & head placed inside the  \\
  &                     & feeding or drinking trough \\ \hline
6 & Nosing floor        & exploratory or manipulative \\
  &                     & behaviour directed at the \\
  &                     & horizontal surface of the \\
  &                     & floor or at the substrate thereon   \\ \hline
\end{tabular}
\caption{The ethogram used to record the gilts' preparturient activity
}
\end{table} 
\pagebreak

\begin{figure}[hbt] 
\centerline{\psfig{figure=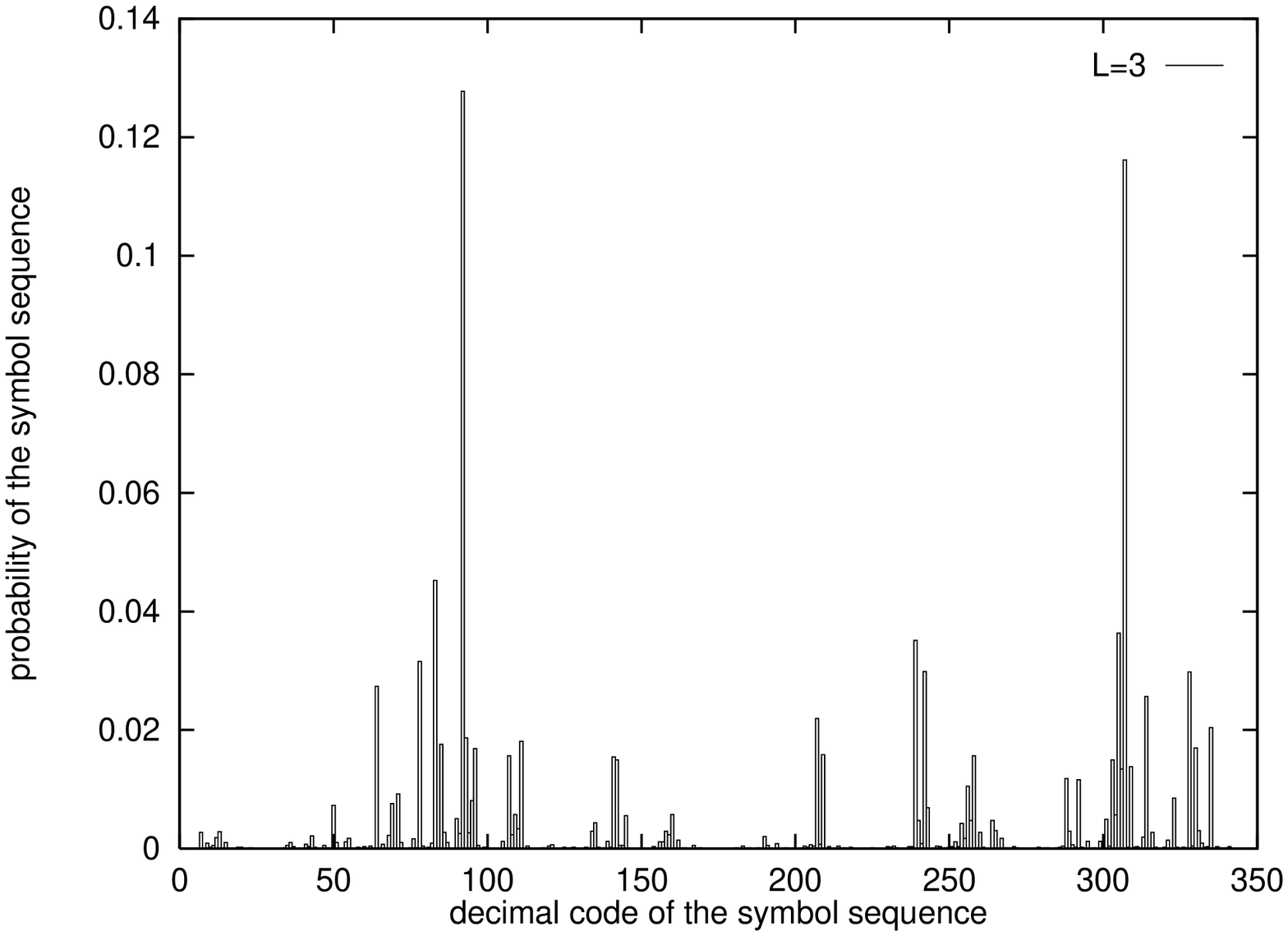,height=10cm}}
\centerline{\psfig{figure=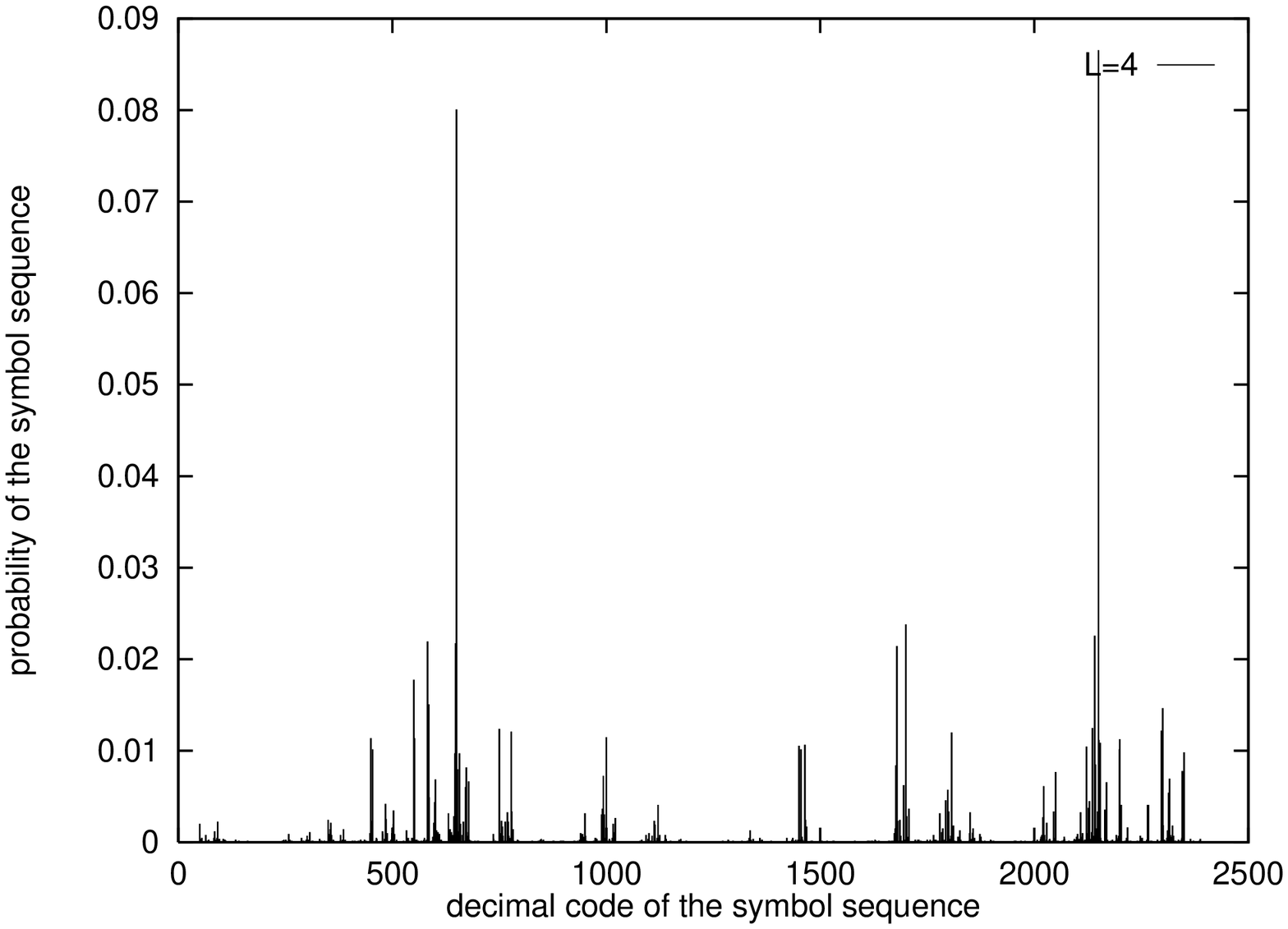,height=10cm}}
\centerline{\psfig{figure=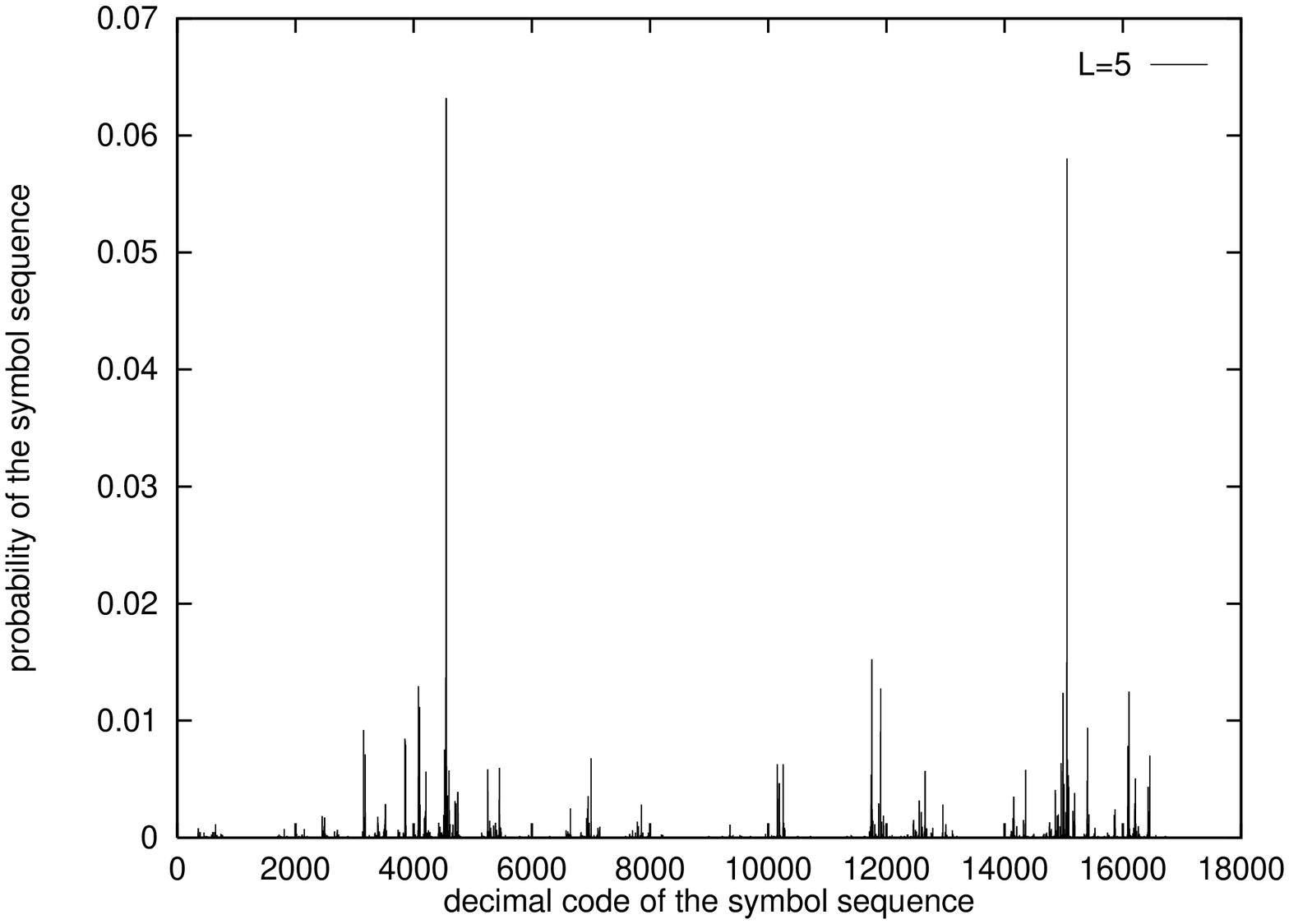,height=10cm}}
\caption{Histogram of the probabilities of the symbolic sequences 
coding behavioural patterns. The video recorded animal activity has 
been translated into a code sequence of codes 0,1,...,6 using the  
classification scheme of Table I. Then the sequential occurrence 
of length 1,2,3,4 and 5 code combinations of symbols had been 
analyzed. The symbol sequences are coded according to the value 
of the symbol sequence as a base 7 number. For example the length 4  
code sequence 6510 is represented by the number $\underline{0}*1+\underline{1}*7+\underline{5}*7^2+ 
\underline{6}*7^3=2310$ In this representation similar sequences get 
close to each other. This number is given on the horizontal axis and 
the corresponding probability is on the vertical axis and the length 
$L=3,4$ and $5$ are shown on {\bf a},{\bf b} and {\bf c} respectively. 
We can see that the probability of high probability sequences changes 
very slowly with $L$, while the rest of the probability is scattered  
among the increasing number of possible symbols.} 
\end{figure} 
 
\pagebreak
\begin{figure}[htb]
\centerline{\psfig{figure=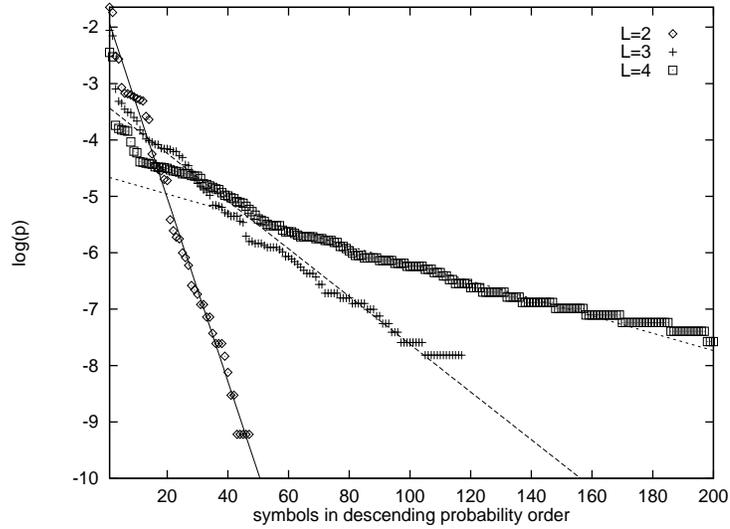,height=10cm}} 
\caption{The same histogram as on Fig. 1, however the symbols are now 
ordered according descending probability on the horizontal axis and 
on the vertical axis the logarithm of the symbol sequence probability 
is shown. The tail of the histogram can be fitted reasonably with 
an exponential. This is the typical behaviour of random symbol 
sequences generated by Markov processes. The head of the histogram 
remains almost in the same position indicating a highly correlated 
behaviour. These high probability sequences are codings of the  
periodic sequence "alert - nosing floor - alert -..." and its variants with 
"rest". These sequences are forced by the nest building instinct of 
the animals. } 
\end{figure} 
 
\pagebreak
\begin{figure}[htb] 
\centerline{\psfig{figure=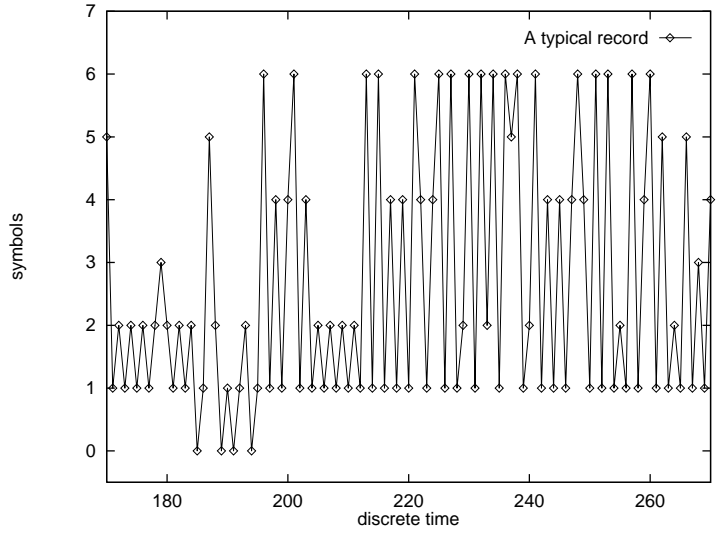,height=10cm}}
\caption{A typical sequence from the time series. The horizontal axis 
is the (discrete) time while the symbols 0 ... 6 are on the 
vertical axis. One can see clearly, that the "1-6-1-..." = 
"alert - nosing floor - alert -..." sequence dominates and it is 
interrupted eventually by other sequences. The random-like changes 
between the regular "1-6-1-6-..." parts are very similar to 
an intermittent time series typical in turbulence.} 
\end{figure} 
 
\pagebreak
\begin{figure}[htb] 
\centerline{\psfig{figure=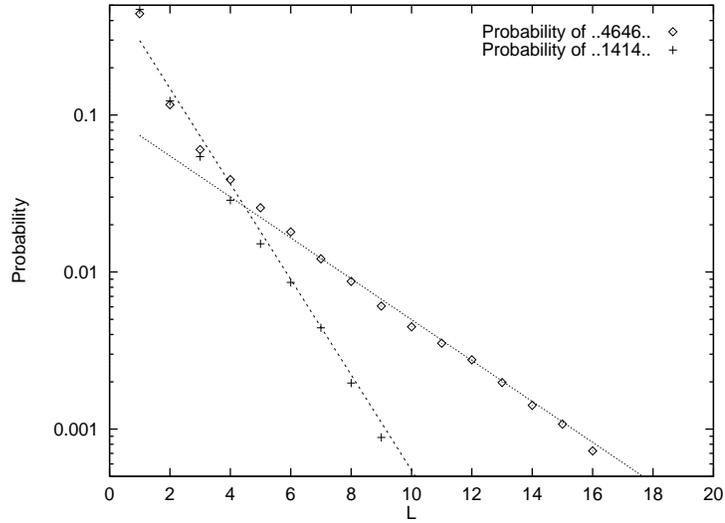,height=10cm}}
\centerline{\psfig{figure=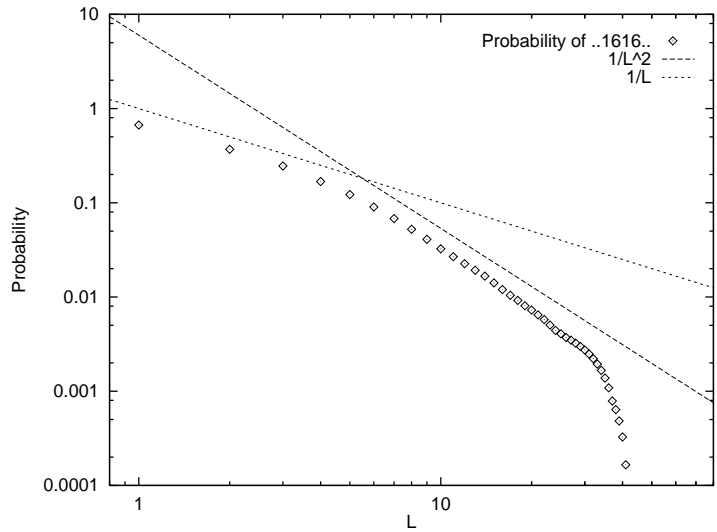,height=10cm}}
\caption{Probability of the periodic sequences "1-4-1-4-..." and
"4-6-4-6-..." {\bf a,}, 
and the probability of the periodic sequence "1-6-1-..." {\bf b,} 
as a function of the length $L$ of the sequence.
One can see that the probability of periodic sequences
decays exponentially with the length except for the "alert - nosing floor - alert -..."
sequence. The $1/L^2$ power law decay is in close analogy with
turbulent flows, where the probability of regular velocity patterns
scales in a similar way. 
}  
\end{figure}

\pagebreak
\begin{figure}[htb] 
\centerline{\psfig{figure=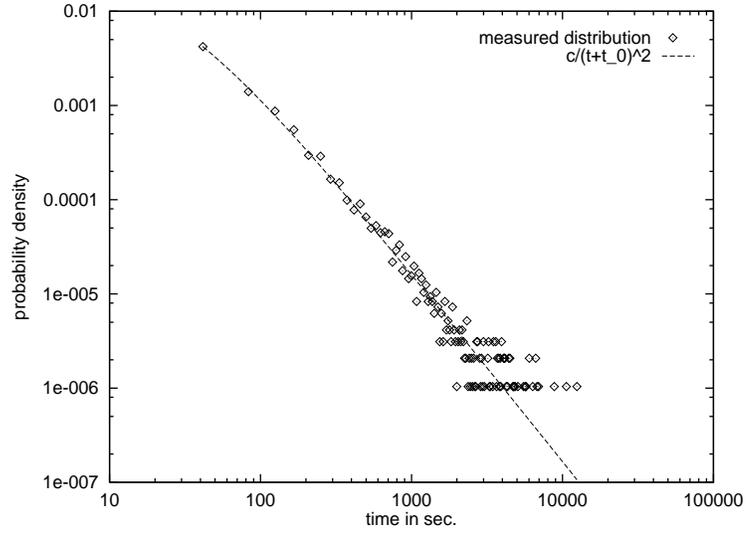,height=10cm}}
\caption{Probability distribution of times spent in a given activity. 
The whole distribution can be fitted with the formula (1), 
where $t_0=21.3\pm0.6$ and $C=16.6\pm0.3$. The limits of validity 
of this formula are beyond the dataset available. The scaling 
behaviour of this distribution is a direct consequence of the  
neuro-hormonal behaviour of the animal. It is not a consequence of 
some scaling  
factors in the local environment of the animal.}  
\end{figure} 
 
\end{document}